\documentclass[a4paper,10pt,twoside]{cpc-hepnp}

\usepackage{multicol}
\usepackage{graphicx}
\usepackage{booktabs}
\usepackage{amssymb,bm,mathrsfs,bbm,amscd}
\usepackage[tbtags]{amsmath}
\usepackage{lastpage}

\begin{document}

\fancyhead[R]{\footnotesize Submitted to 'Chinese Physics C'}

\footnotetext[0]{Received 10 Oct 2013}

\title{An analysis on optimization of undulator in self-seeding free electron laser\thanks{partly supported by the Major State Basic Research Development Program of China (2011CB808301) and National Natural Science Foundation of China (11375199) }}

\author{%
      JIA Qi-Ka(贾启卡)
} \maketitle

\address{%
National Synchrotron Radiation Laboratory, University of Science and
Technology of China, Hefei, 230029, Anhui, China\\
}

\begin{abstract}
A simple analysis is given for optimum length of undulator in self-seeding free electron laser (FEL). The obtained relations show the correlation between the undulator length and the system parameters. The power required for the seeding in the second part undulator and overall efficiency to monochromatizating the seeding settle on the length of the first part undulator; the magnitude of seeding power dominates the length of the second part undulator; the whole length of the undulators in self-seeding FEL is determined by the overall efficiency to get coherent seed, it is about half as long again as that of SASE, not including the dispersion section. The requirement of the dispersion section strength is also analyzed.
\end{abstract}



\footnotetext[0]{\hspace*{-3mm}\raisebox{0.3ex}{$\scriptstyle\copyright$}2013
Chinese Physical Society and the Institute of High Energy Physics
of the Chinese Academy of Sciences and the Institute
of Modern Physics of the Chinese Academy of Sciences and IOP Publishing Ltd}%

\begin{multicols}{2}

\section{Introduction}

The self-amplified spontaneous emission (SASE) free-electron lasers (FELs) in the x-ray region of the spectrum are currently opening up new frontiers across science. Although SASE FELs have brightness up to 108 times greater than laboratory sources their full potential is limited by a relatively poor temporal coherence. To improve the temporal coherence of SASE is an important research hot sport, many scheme are proposed. One attractive scheme is self-seeding FEL, the idea was proposed at DESY to generate a narrow spectrum soft and hard X-FEL [1,2]. In the scheme, the undulators is divided into two parts by a magnet chicane. The first part of the undulator is operated in the SASE linear regime, the output of the radiation pass a monochromator to generate a narrow spectrum seeding laser, and is input into the second part of the undulator, in which it is amplified to saturation. The magnetic chicane delays the electron bunch to match the light delay caused by monochromator. In these schemes a long (gentle) chicane is needed to depress electron energy spread due to synchrotron radiation, that is not convenient. The improved schemes were proposed to avoid this long chicane, one of them is use two separate electron bunches[3], in which the delayed seed from the first electron bunch will be amplified by the second electron bunch; A new approach of monochromatization for hard X-FEL is proposed [4] at DESY again, so that a compact magnetic chicane can be adopted. Using this approach the self-seeding X-FEL experiment has been successfully demonstrated at SLAC laboratory [5], the results are encouraging, and the bandwidth of X-FEL is reduced significantly.

The basic conditions for self-seeding FEL are analyzed in Ref [1]. In this paper, we give more detail analysis on the optimization of undulator length in it. A disadvantage of the self-seeding FEL is that total length of the undulator system is longer than that of SASE scheme. For self-seeding FEL scheme to get the optimum performance and make the system length as short as possible, it is crucial to optimize the length of undulators and determine where the monochromator and the electron by-pass chicane should be insert.

\section{Analysis}

The first part of the undulator is operated in the exponential gain regime of SASE, but not to saturation regime, correspondingly the length of it should be long enough, but shorter than the saturation length. The typical saturation length of SASE is about twenty gain length, so it has ~$L_1  < L_s \sim20L_g$~. The out power of the first part of the undulator is~$P_1  = P_{ef} e^{L_{\rm{1}} /L_g } /9$~, where~$P_{ef}$~ is the effective SASE start up power (equivalent starting noise power). After monochromatizating, the input power of the second undulator i.e. the self-seeding power is
\begin{eqnarray}
P_{20}  = P_1 \eta  = \frac{{\rm{1}}}{9}P_{ef} e^{\frac{{L_{\rm{1}} }}{{L_g }}} \eta
\end{eqnarray}
where~$\eta  = \eta _m  \times b$~: ~$\eta _m$~is transmission efficiency of the monochromator，~$\times b$~is the fraction of total power that in the required narrow bandwidth: ~$b = \Delta \lambda _m /\Delta \lambda _1 $~. ~$\Delta \lambda _1$~is the bandwidth of SASE at the exit of the first part of the undulator：~$\Delta \lambda _1 (rms) \simeq \lambda \sqrt {\rho /N_1 }$~, ~$\rho$~is FEL parameter, ~$N_1$~ is periods numbers of the first part of the undulator, ~$\Delta \lambda _m$~is the resolution of the monochromator，it should satisfy the condition [1]:
\begin{eqnarray}
\lambda /\pi \sigma _z  \le \Delta \lambda _m /\lambda  <  < \Delta \lambda _1 /\lambda
\end{eqnarray}
where ~$\sigma _z$~is the length of electron bunch. Due to noise start-up, SASE is chaotic light with ~$M_L$~ spikes in spectral profile on average, while the average bandwidth of spikes is about ~$\lambda /\pi \sigma _z$~, namely ~$M_L  = \lambda /\pi \sigma _z /\Delta \lambda _1  = \sqrt {N_1 /\rho } /\pi \sigma _z $~.  Therefore we have
\begin{eqnarray}
1/M_L  \le b <  < 1
\end{eqnarray}

The self-seeding power should be much larger than the effective SASE start up power ~$P_{ef}$~. We denote the ratio of the self-seeding power to the effective SASE start up power to be ~$\alpha$~, namely we want ~$\alpha  = P_{20} /P_{ef}  = \eta e^{L_{\rm{1}} /L_g } /9 >  > {\rm{1}} $~. Thus the length of the first part of the undulator should be
\begin{eqnarray}
L_{\rm{1}} {\rm{ = (}}\ln \frac{{9\alpha }}{\eta })L_g  >  > \ln (\frac{9}{\eta })L_g
\end{eqnarray}

\begin{center}
\includegraphics[width=8.0cm]{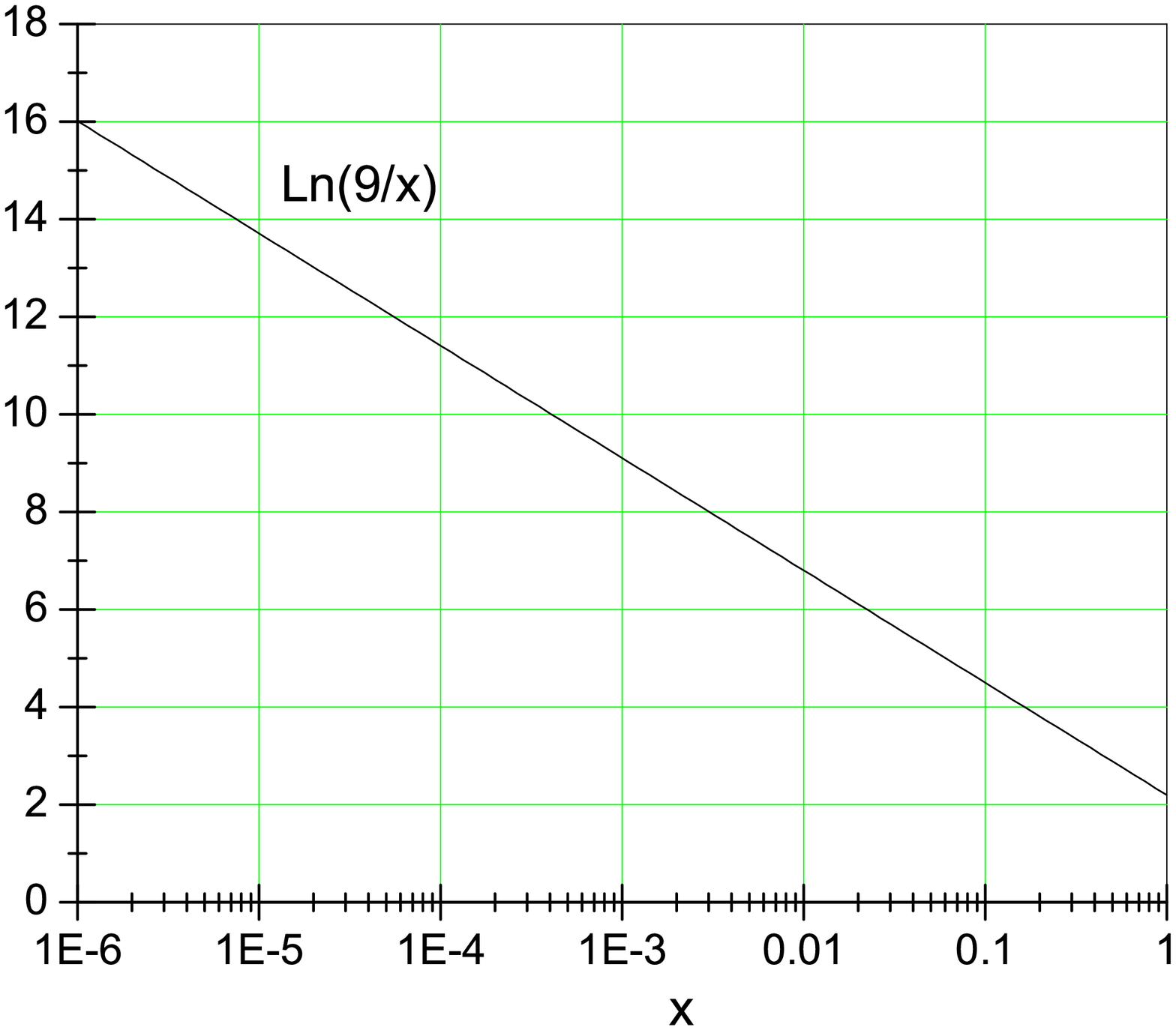}
\figcaption{\label{fig1} The logarithmic relationship on length of the undulator(Eq.(4) and Eq.(10)).}
\end{center}

For an example, ~$\eta _m  \sim 10^{ - 1}  - 10^{ - 2} $~, ~$b >  \sim 10^{ - 2}$~, namely ~$\eta \sim 10^{ - 3}  - 10^{ - 4}$~, if we require ~$\alpha  \sim 10  - 10^{ 2}$~, then we have  ~$L_1  \sim (12 - 16)L_g$~(Fig.1). The longer length of the first part undulator can provide the higher seeding power, but it induces the larger energy spread that will suppress the FEL gain. The electron energy modulation induced by SASE in the first part undulator can be given from the electron energy equation
\begin{eqnarray}
\frac{{d\gamma ^2 }}{{d{\rm{z}}}} \simeq  - 2a_u [JJ]k_s {\mathop{\rm Re}\nolimits} (\tilde a_{s1} e^{i\phi } )
\end{eqnarray}
where ~$\tilde a_s  = a_s e^{i\varphi _s }$~,~$a_s  = eE_s /mc^2 k_s$~ and ~$a_u  = eB_u /mc^2 k_u$~ are dimensionless vector potential of the ~$rms$~ radiation field ~$E_s$~and undulator field ~$B_u$~, respectively;~$k_s  = 2\pi /\lambda _s$~and~$k_u  = 2\pi /\lambda _u$~are the corresponding wave number;~$\left[ {JJ} \right]$~is the Bessel function factor. With~$\tilde a_{s1}  \approx a_{ef} e^{\mu _1 z} /3, _{} \mu _1  = k_u \rho (i + \sqrt 3 )$~, ~$a_{ef}$~is the effective star up shot noise field, we get the maximum electron energy modulation induced by SASE in exponential gain regime:
\begin{eqnarray}
\Delta \gamma _m /\gamma  = \frac{{k_s a_{s1} a_u [JJ]}}{{\gamma ^2 }}\sqrt 3 L_g {\rm{ = }}2\rho \sqrt {P_1 /\rho P_e }
\end{eqnarray}
It should be smaller than ~$\rho$~, then for the power of the first part of the undulator it has
\begin{eqnarray}
P_1  < \rho P_e /4
\end{eqnarray}
For the second part of the undulator, the optical power should reach saturation：
\begin{eqnarray}
\frac{{\rm{1}}}{9}P_{20} e^{L_{\rm{2}} /L_g } {\rm{ = }}P_s {\rm{ }}
\end{eqnarray}
Where the saturation power~$P_s$~is same as SASE's: ~$P_s  = P_{ef} e^{L_{\rm{s}} /L_g } /9$~,~$L_s$~is the saturation length of SASE. Substituting the expression of~$P_s$~ and ~$P_{20}  = \alpha P_{ef}$~into Eq.(8), we give the length of the second part of the undulator as
\begin{eqnarray}
L_{\rm{2}}  = L_s  - (\ln \alpha )L_g
\end{eqnarray}
Thus the length of the second part of the undulator is dominated by the magnitude of seeding power. From Eq.(4) and (9), the total length of the undulator is
\begin{eqnarray}
L_{\rm{1}}  + L_{\rm{2}}  = L_s {\rm{ + (}}\ln \frac{9}{\eta })L_g
\end{eqnarray}
Therefore the whole length of the undulators in self-seeding FEL is determined by the overall efficiency to get coherent seed, it is longer than that of SASE case. For example,~$\eta \sim 10^{ - 3}  - 10^{ - 4}$~, ~$\Delta L \sim (9 - 12)L_g$~(Fig.1), is about half as long again as that of SASE, not including the magnetic chicane section.
\begin{center}
\includegraphics[width=8.0cm]{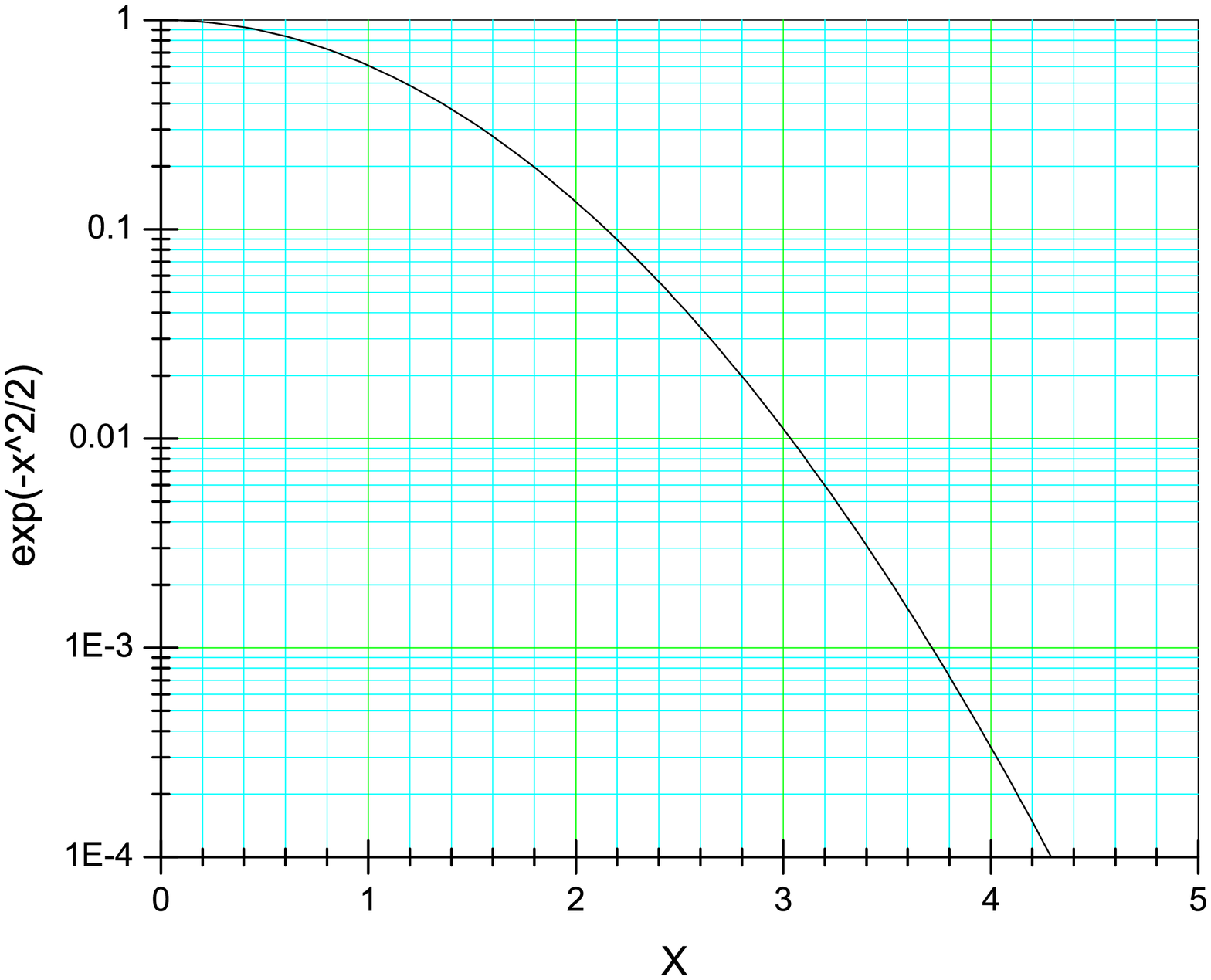}
\figcaption{\label{fig2} The exponential term in ~$b_{20}$~.}
\end{center}

Now we consider the requirement of the dispersion intensity of the magnetic chicane. Besides delaying the electron bunch and diverting the electron beam around the monochromator. The dispersion field of magnetic chicane must wash out the electron microbunching induced by SASE in the first part of the undulator. The bunching factor at the end of the first undulator is [6]
\begin{eqnarray}
b_1  = \frac{{b_{10} }}{3}e^{{{L_1 } \mathord{\left/{\vphantom {{L_1 } {2L_g }}} \right.\kern-\nulldelimiterspace} {2L_g }}}  = \sqrt {\frac{{P_1 }}{{\rho P_e }}}
\end{eqnarray}
where ~$b_{10}$~is the initial bunching factor of SASE. After the chicane and at the entrance of the second undulator it becomes
\begin{eqnarray}
b_{20}  \simeq J_1 (D\frac{{\Delta \gamma _m }}{\gamma })\exp [ - \frac{1}{2}(D\frac{{\sigma _\gamma  }}{\gamma })^2 ]
\end{eqnarray}
where ~$D \simeq 4\pi (N_d  + N_1 /2)$~,~$N_1$~is the periods number of the first part of the undulator,~$N_d$~is the dispersive parameter of the dispersion section; and ~$\Delta \gamma _m /\gamma$~is the electron beam energy modulation induced by SASE in the first part of the undulator and given by Eq.(6). The bunching factor at the entrance of the second undulator (Eq.(11)) must not be larger than the initial bunching factor of SASE, which typically is about ~$10^{ - 4}  - 10^{ - 3}$. Because the maximum of first order Bessel function ~$J_1  < 0.6$~, thus the exponential term in the right hand of Eq.(11) should have ~$\exp [ - (D\sigma _\gamma  /\gamma )^2 /2] \le 10^{ - 3}$~, so one can take~$D > \sim 4\gamma /\sigma _\gamma$~(Fig.2), namely ~$R_{56}  > \sim{{2\gamma \lambda _s } \mathord{\left/{\vphantom {{2\gamma \lambda _s } {\pi \sigma _\gamma  }}} \right.\kern-\nulldelimiterspace} {\pi \sigma _\gamma  }}$~. In practical design of the magnetic chicane, an appropriate electron trajectory offset and delay to match the seed laser delay also should be considered.

\section{Summary}
In summary, the optimum length of undulator in self-seeding FEL is analyzed. The relations between the undulators length and the system parameters are obtained. The power required for the seeding in the second part undulator and overall efficiency to monochromatizating the seeding settle on the length of the first part undulator; the length of the second part undulator are dominated by the magnitude of seeding power; and the whole length of the undulators in self-seeding FEL is determined by the overall efficiency to get coherent seed, typically is about half as long again as that of SASE, not including the dispersion section. The needed strength of the dispersion section is also analyzed. The works here can help design and optimization of self-seeding FEL. The comparisons with results of the simulation or the experiments are needed in following works.

\end{multicols}

\vspace{-1mm}
\centerline{\rule{80mm}{0.1pt}}
\vspace{2mm}

\begin{multicols}{2}

\end{multicols}
\clearpage

\end{document}